# Thermodynamic conditions governing the optical temperature and chemical potential in nonlinear highly multimoded photonic systems

MIDYA PARTO[1], FAN O. WU[1], PAWEL S. JUNG[1,2], KONSTANTINOS MAKRIS[3], DEMETRIOS N. CHRISTODOULIDES[1,*]

[1]*CREOL, College of Optics and Photonics, University of Central Florida, Orlando, Florida 32816-2700, USA*
[2]*Faculty of Physics, Warsaw University of Technology, Warsaw, Poland*
[3]*Department of Physics, University of Crete, 71003 Heraklion, Greece, EU*
*\*Corresponding author: demetri@creol.ucf.edu*

**We show that, in general, any complex weakly nonlinear highly multimode system can reach thermodynamic equilibrium that is characterized by a unique temperature and chemical potential. The conditions leading to either positive or negative temperatures are explicitly obtained in terms of the linear spectrum of the system, the input power, and the corresponding Hamiltonian invariant. Pertinent examples illustrating these results are provided in various scenarios.**

Multimode optical systems are ubiquitous in photonics. Such many-mode structures are nowadays used in generating frequency combs [1] and for implementing fiber-based lasers [2] and spatial mode-multiplexed high-capacity communication systems [3, 4]. One of the advantages offered by such arrangements is their capability in handling high power levels, in which case, light dynamics tend to exhibit nonlinear behaviors [5]. In this regime, the synergy between nonlinear effects and multimode interactions can lead to an array of novel optical effects that are otherwise impossible in single-mode optical platforms. These include for example, optical microcavity solitons [6], spatiotemporal mode-locking [2], new Cherenkov dispersive wave lines and supercontinuum generation [7-9], as well as geometric parametric instabilities [8, 10-12], to mention a few. Lately, a rather peculiar effect has been consistently reported in several experimental studies whereby the power in a multimode nonlinear fiber was found to eventually settle into the lower group of modes: the so-called beam self-cleaning effect. This intriguing mechanism, has nothing to do with any Raman and/or self-focusing effects, instead is a byproduct of multi-wave mixing energy exchange [8, 13, 14]. Quite recently, a systematic explanation was provided in interpreting this effect based on thermodynamic notions [15-17]. In this formalism, the complex interactions in these heavily multimoded systems are treated via statistical mechanics, and the beam self-cleaning effect is explained as a thermalization process aiming to extremize the underlying optical entropy. In such a setting, the power distribution is governed by an optical temperature and a corresponding chemical potential. At this point, the question naturally arises as to whether these thermodynamic parameters are unique, and if so, what are the general conditions determining their accessible range.

In this letter, we formally show that the resulting temperature and chemical potential in these highly multimoded nonlinear systems are indeed unique. For a given multimode structure with a specific linear spectrum (propagation constants or eigenfrequencies), these two intensive parameters can be directly determined as a function of the input power and the initial excitation conditions. Our analysis can explicitly provide the critical points that mark the transition between positive and negative temperatures, while at the same time can indicate their allowed domain. These analytical results are further augmented with numerical simulations carried out in various multimode nonlinear arrangements.

We begin our analysis by considering an arbitrary nonlinear multimode optical system supporting a finite number of $M$ discrete states. This configuration can be treated either electrodynamically (e.g. multimode waveguides or cavities) or on occasions discretely via coupled mode theory (like waveguide arrays or coupled-resonator-optical-waveguides (CROWs)) [18, 19]. Under linear conditions, the eigenmodes of this system form an orthonormal set $|\psi_i\rangle$ with real, normalized eigenvalues $\mathcal{E}_i$, satisfying $H_L|\psi_i\rangle = \mathcal{E}_i|\psi_i\rangle$, where $H_L$ represents the corresponding optical Hamiltonian operator. Without any loss of generality, we assume that these eigenvalues can be progressively arranged according to $\mathcal{E}_1 \leq \mathcal{E}_2 \leq \cdots \leq \mathcal{E}_M$, where $\mathcal{E}_1$ denotes the eigenvalue of the highest-order mode, while $\mathcal{E}_M$ corresponds to that associated with the ground state. The equalities in this latter equation indicate possible degeneracies in the spectrum. This particular sorting applies in multimode waveguide structures, whereas it has to be reversed in optical cavity arrangements. The power levels in this multimode configurations are here taken to be relatively low, in which case, the expectation value $U \equiv -\langle\Psi|H_L|\Psi\rangle$ of the dominating linear optical Hamiltonian remains invariant during evolution, and represents the "internal energy" of the system, i.e. $U = -\sum_{i=1}^{M} \mathcal{E}_i |c_i|^2$. In this latter equation, the coefficients $|c_i|^2$ denote the average value of the normalized power in each mode, while the negative sign stems from the fact that the spectrum is inverted in waveguide arrangements. Note that this constant internal energy is prespecified by the initial power launching distributions $|c_{i0}|^2$ at the input, in other words $U = -\sum_{i=1}^{M} \mathcal{E}_i |c_{i0}|^2$. In this closed arrangement, another conserved quantity is the normalized power $\mathcal{P}$ conveyed in the system, i.e. $\mathcal{P} =$

$\sum_{i=1}^{M}|c_i|^2$. In these arrangements, the thermodynamically extensive variables $(U, M, \mathcal{P})$ are related to each other through the optical temperature $T$ and chemical potential $\mu$ associated with the system. This relation is given through the following equation of state that explicitly involves the number of modes, $U - \mu\mathcal{P} = MT$ [15]. In general, the system is expected to reach thermal equilibrium by maximizing its entropy. In this case, the thermalized average power levels conveyed by each mode are found to positive. The equivalent temperature $T$ can now be obtained by determining the roots of the function $f(x)$. As we will see, the thermodynamically relevant roots that keep $|c_i|^2 > 0$ should reside outside the interval $(\rho_1, \rho_N)$. This can be directly shown by noting

$$|c_i|^2 = -\frac{\mathcal{P}T}{\rho_i - x} > 0. \quad (3)$$

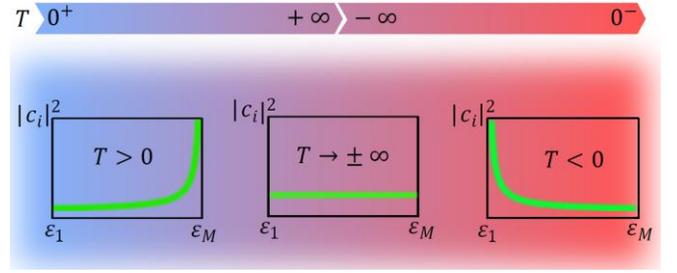

**Fig. 2.** Different regimes of positive and negative temperatures associated with various equilibrium modal occupancies in a highly-multimode nonlinear optical arrangement.

From here, one concludes that $(\rho_i - x)$ should not change sign as $i$ progresses from 1 to $N$. This can only be true if indeed $x < \rho_1$ or $x > \rho_N$. In what follows, we will show that under these conditions, the equivalent temperature $T$ is in fact unique for a given set of $(U, M, \mathcal{P}, \{\mathcal{E}_i\})$. It can be directly proved from Eq. (2) that $f(x)$ can be cast in an equivalent polynomial of degree $N - 1$, thus having $N - 1$ roots. Interestingly, these $N - 1$ roots are all real, but only one of them is physical, since it falls outside the range $(\rho_1, \rho_N)$. To illustrate the principle behind this proof, let us examine the behavior of $f(x)$ for two different cases (with and without degeneracy) as shown in Fig. 1. Note that $U$ lies in the interval $-\mathcal{P}\mathcal{E}_M < U < -\mathcal{P}\mathcal{E}_1$. As a result, $\rho_1 < \cdots \rho_k < 0 < \rho_{k+1} \cdots < \rho_N$. In other words, somewhere in the sequence, the signs associated with $\rho_i$ switch from negative to positive, depending on the value of $U$. This in turn, influences the branches of $f(x)$ around the $\rho_i$ singularities as shown in Fig. 1. In the range where successive $\rho_k$ are all positive (or negative), the $f(x)$ function transitions from positive to negative values (or from negative to positive) between $\rho_k$ and $\rho_{k+1}$. From this change in sign in these $N - 2$ compartments, Bolzano's theorem guarantees $N - 2$ real roots [21]. We next consider the asymptotic behavior of the $f(x)$ function at $x \to \pm\infty$

$$\lim_{x\to\pm\infty} f(x) = \sum_{i=1}^{N}\frac{g_i}{1 - (\rho_i/x)} - M = \frac{1}{x}\sum_{i=1}^{N} g_i\rho_i. \quad (4)$$

Note that immediately outside the $\rho_k$ interval, i.e. at $x = \rho_N + \epsilon$ and $x = \rho_1 - \epsilon$ ($\epsilon > 0$), the function $f(x) > 0$. On the other hand, Eq. (4) indicates that $f(x)$ becomes negative for $x \to +\infty$ (in the right open interval) whenever $\sum_{i=1}^{N} g_i\rho_i < 0$ or for $x \to -\infty$ (in the left open interval) provided that $\sum_{i=1}^{N} g_i\rho_i > 0$. This change in sign guarantees an additional real root in either of these two open intervals. This brings the number of real roots to be at least $N - 1$, which is actually what is expected from $N - 1$ roots of the polynomial $f(x)$. This completes the proof. Indeed, $f(x)$ has $N - 1$ real roots out of which only one is thermodynamically relevant (so as all $|c_i|^2 > 0$). This physically relevant root always lies in one of the two open intervals and hence uniquely assigns a temperature $T$ to the system. If this root $x' > \rho_N$ the temperature is positive, otherwise if the root is $x' < \rho_1$ the temperature is negative. This is consistent with the behavior of the examples depicted in Figs. 1(a, b).

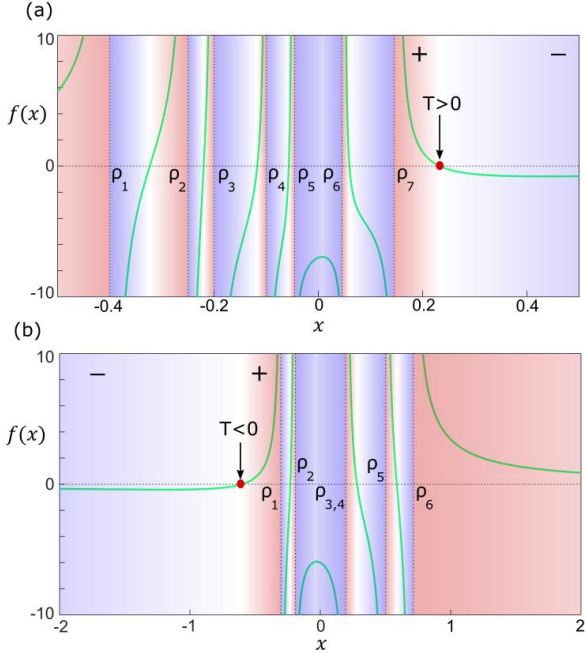

**Fig. 1.** The thermodynamically meaningful root (red point) of the function $f(x)$ in two different cases when the linear spectrum of the system (a) is not degenerate and (b) when is degenerate. For these two examples, the values of $\rho_k$'s are such that lead to either a positive or negative optical temperature $T$.

obey a Rayleigh-Jeans distribution, i.e. $|c_i|^2 = -T/(\mathcal{E}_i + \mu)$ [15-17, 20]. From here, one can readily deduce that

$$\mathcal{P} = -\sum_{i=1}^{M}\frac{T}{\mathcal{E}_i + \frac{U - MT}{\mathcal{P}}}. \quad (1)$$

Equation (1) clearly indicates that if the linear spectrum $\mathcal{E}_i$ of the structure is known, along with the input power $\mathcal{P}$ and internal energy $U$, one should then be able to obtain the respective temperature, and consequently, the chemical potential $\mu$. As it will turn out, this latter algebraic Eq. (1) possesses only real roots. Yet, it is still not clear whether a temperature $T$ uniquely exists in such a way that the power occupancies $|c_i|^2$ are positive for all modes. To address this problem, we rewrite Eq. (1) as

$$f(x) = \sum_{i=1}^{N} g_i \frac{x}{x - \rho_i} - M = 0, \quad (2)$$

where $x = MT$, $\rho_i = U + \mathcal{P}\mathcal{E}_i$ and $g_i$ stands for the degree of degeneracy (if any) associated with a particular mode $|\psi_i\rangle$. In this same representation, $N$ denotes the number of distinct eigenvalue levels, and thus $\sum_{i=1}^{N} g_i = M$. It is important to emphasize that the quantities $\rho_i$ are also progressively sorted according to $\rho_1 < \rho_2 < \cdots < \rho_N$, since the power $\mathcal{P}$ is always

Table.1 Conditions governing positive/negative temperatures

| Temperature | $\bar{\mathcal{E}}$ –average spectrum relation | Temperature domain |
|---|---|---|
| positive: $T > 0$ | $-\dfrac{U}{\mathcal{P}} > \bar{\mathcal{E}} = \dfrac{1}{M}\sum_{i=1}^{N} g_i \mathcal{E}_i$ | $T > \dfrac{U + \mathcal{P}\mathcal{E}_N}{M}$ |
| negative: $T < 0$ | $-\dfrac{U}{\mathcal{P}} < \bar{\mathcal{E}} = \dfrac{1}{M}\sum_{i=1}^{N} g_i \mathcal{E}_i$ | $T < \dfrac{U + \mathcal{P}\mathcal{E}_1}{M}$ |

Table 1 summarize our results. Here $\bar{\mathcal{E}}$ denotes a transition point representing the average value of the spectrum. These relations provide the conditions for positive and negative temperatures and indicate in each case the range where the physically relevant real root can be obtained. If in other hand the quantity $-U/\mathcal{P}$ approaches from above or below the transition point $\bar{\mathcal{E}}$, then the resulting temperature tends to either $T \to +\infty$ or $-\infty$. Finally, as $U \to -\mathcal{P}\mathcal{E}_M$ all the power condensates in the ground state (condensate) and thus $T \to 0^+$. The opposite is true as $U \to -\mathcal{P}\mathcal{E}_1$ where the power condensates into the highest-order mode(s), a phenomenon expected at $T \to 0^-$.

This temperature scale is shown schematically in Fig. 2. In this scale negative temperatures are "hotter" than hot [22]-a relation that dictates any resulting energy flow $\Delta U$. While for positive temperatures the lower group of modes is mostly occupied (since $|c_i|^2 = -T/(\mathcal{E}_i + \mu)$) as in the case of beam self-cleaning [8, 13, 14], the opposite is true for negative temperatures where the higher-order modes are populated (Fig. 2). On the other hand, if the temperature is infinite all modes are equally loaded. Once the temperature is determined from Eq.(2) for a given set of $(U, M, \mathcal{P}, \{\mathcal{E}_i\})$, the corresponding chemical potential can be obtained from the equation of state, i.e. $\mu = (U - MT)/\mathcal{P}$. We will now illustrate these results through pertinent examples.

To do so, we consider an optical, Kerr nonlinear multimode system in the form of a multicore arrangement. Each core is assumed to be single-moded and hence in the tight-binding approximation the array has one band. For simplicity, we assume a square $L \times L$ lattice. The normalized evolution equation (along the propagation axis $z$) governing the nonlinear interactions in this array is given by:

$$i\frac{dA_{m,n}}{dz} + \kappa_1\big(A_{m-1,n} + A_{m+1,n}\big) + \kappa_2\big(A_{m,n-1} + A_{m,n+1}\big) + |A_{m,n}|^2 A_{m,n} = 0, \quad (5)$$

where, $A_{m,n}$ represent the optical modal field amplitude at the waveguide site $(m,n)$ and $\kappa_1$ and $\kappa_2$ are the coupling coefficients between waveguides in the two transverse directions. In this system, the initial mode occupancies at the input can be obtained from $|c_{i0}|^2 = |\langle \psi_i | \Psi \rangle|^2$, where $\vec{\Psi} = (A_{1,1}, A_{1,2}, \ldots A_{L,L})^T$ denotes the input complex optical field vector and $|\psi_i\rangle$ represent a supermode eigenvector in this multimode arrangement. In this discrete system, the linear ordered set of eigenvalues can be readily obtained from

$$\mathcal{E}_i = 2\kappa_1 \cos\left(\frac{k\pi}{L+1}\right) + 2\kappa_2 \cos\left(\frac{l\pi}{L+1}\right) \quad (6)$$

where the integers $(k, l)$ take values from the set $\{1,2,..,L\}$. In addition the linear part of the total Hamiltonian, that is assumed to dominate the process in this quasi-linear array, is given by the conserved quantity $U = -\sum_{i=1}^{M} \mathcal{E}_i |c_i|^2$. Given that $U, \mathcal{P}$ are invariants, one can evaluate them from the initial conditions by setting $c_i \to c_{i0}$. On the other hand the instantaneous coefficients $c_i(z)$ can be obtained through projections on the supermodes $|\psi_i\rangle$. Throughout evolution, the ensemble averages $\langle |c_i|^2 \rangle$, simply denoted here as $|c_i|^2$, are monitored as a function of $z$. Under the action of a weak nonlinearity, at thermal equilibrium, the average modal occupancies settle into a Rayleigh-Jeans distribution, $|c_i|^2 = -T/(\mathcal{E}_i + \mu)$.

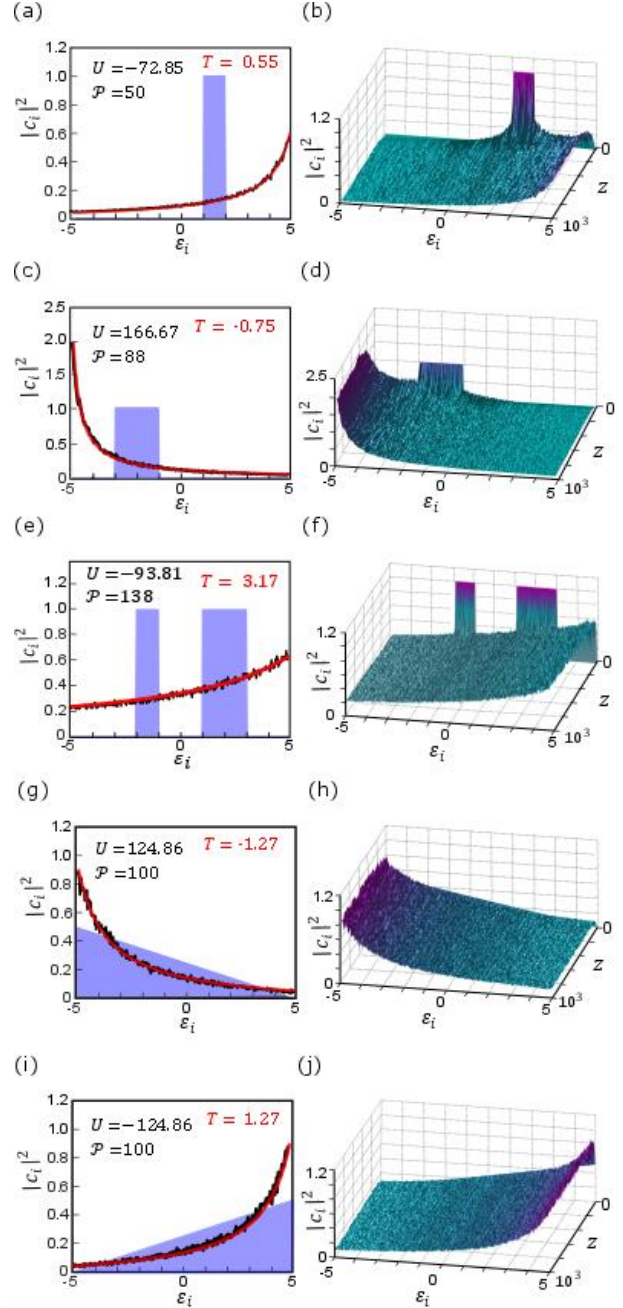

**Fig. 3.** Several examples illustrating nonlinear multimode evolution in a square array of $20 \times 20$ optical waveguides having two transverse coupling constants $\kappa_1 = 1$ and $\kappa_2 = 1.5$. The spectrum of this array is obtained from Eq. (6) and ranges approximately from $-5$ to $5$. In each case, the left panel indicates the initial $|c_{i0}|^2$ (shown in blue) and final (black and red plots) mode occupancy distributions, as well as the equilibrium optical temperature. The right panels depict the evolutions of these occupancies along the propagation direction $z$. The left panels also provide the input energy $U$ and power $\mathcal{P}$ in each case.

For demonstration purposes, let us now consider the following example. We assume that $L = 20$ and hence the square array supports $M = 20 \times 20 = 400$ modes. Moreover, let the system be anisotropic with $\kappa_1 = 1$ and $\kappa_2 = 1.5$. In this case, the spectrum (Eq.(6)) is approximately bounded between $(-5,5)$ with an average value of $\bar{\mathcal{E}} = 0$. In all situations examined here, the normalized length of the system is taken to be $z_{max} = 10^3$. As a first scenario (Fig.3.(a)), the input excitations $|c_{i0}|^2 = 1$ are uniformly distributed within the range $1 \leq \mathcal{E}_i \leq 2$. For these values, $U = -72.85$ and $\mathcal{P} = 50$. The resulting modal distributions $|c_i|^2$ are then obtained by numerically solving Eq.(5) after averaging the results of several ensembles, all initiated with the same $|c_{i0}|$ amplitudes but different random phases ($arg(c_{i0})$). The temperature of the system, after thermalization, can now be theoretically obtained (by solving Eq.(2) in conjunction with Table.1). In doing so, one finds that for this example $T = 0.55$ (hence $= -5.84$). Ensemble averages performed on this same configuration (after numerically solving Eq.(5)), indeed confirm that the resulting distribution is of the Rayleigh-Jeans type with a best fit temperature $T = 0.54$ ($\mu = -5.83$), in excellent agreement with analytical results. The evolution of the corresponding average (ensemble) power distributions $|c_i|^2$ is depicted in Fig.3(b) as a function of distance. A comparison between the theoretically anticipated distribution and that obtained numerically is also provided in Fig.3(a). The same system is now excited with a different initial distribution that favours higher-order modes, for which $|c_{i0}|^2 = 1$ contained in the interval $-3 \leq \mathcal{E}_i \leq -1$. The corresponding $U, \mathcal{P}$ are also provided in Fig.3(c). For this set, one can readily conclude that the system will settle into a negative temperature (Table.1), having a value of $T = -0.75$ after that (from Eq.(2)) and corresponding $\mu = 5.29$. Ensemble averages numerically performed on this latter configuration confirm these predictions. Three more scenarios are provided in Figs.3(e-j). In all occasions there is an excellent agreement between analytical predictions and numerical simulations. In the last two examples (Fig.3(g-j)) the initial energy distributions among modes is monotonically increasing or decreasing with the mode number. In such a case, it is then possible to quickly predict the sign of the resulting thermalization temperature by using the Chebyshev's sum inequalities. For example, if $|c_{10}|^2 \geq |c_{20}|^2 \geq \cdots \geq |c_{M0}|^2$, and since $\mathcal{E}_1 \leq \mathcal{E}_2 \leq \cdots \leq \mathcal{E}_M$, then these inequalities imply that $\sum_{i=1}^{M} \mathcal{E}_i |c_{i0}|^2 \leq M^{-1} \sum_{i=1}^{M} \mathcal{E}_i \sum_{i=1}^{M} |c_{i0}|^2 = \bar{\mathcal{E}}\mathcal{P}$, from where we can deduce (Table.1) that the thermalization temperature is negative $T < 0$. The converse is true if the monotonic relation between input excitations $|c_{i0}|^2$ is reversed. In all the examples presented in Fig.3 we found that the system can reach thermal equilibrium for $z \gtrsim 200$.

In conclusion, we have shown that at thermal equilibrium, the optical temperature and chemical potential associated with a weakly nonlinear highly multimode system can be uniquely determined by the initial excitation parameters $(U, P)$ and the corresponding linear spectrum. The conditions leading to either positive or negative temperatures are also explicitly obtained. We would like to emphasize that our results are general and universally apply to any optical nonlinear multimode system, irrespective of whether the structure is continuous or discrete. These results could be relevant in designing high-power fiber nonlinear sources and heavily multimoded cavities where multi-wave nonlinear interactions play an important role.

## Acknowledgements
This work was supported by the Office of Naval Research (ONR) (MURI N00014-17-1-2588), Office of Naval Research (ONR) (N00014-18-1-2347), National Science Foundation (NSF) (EECS-1711230), and Qatar National Research Fund (QNRF)(NPRP9-020-1-006). P. S. J. thanks the Polish Ministry of Science and Higher Education for Mobility Plus scholarship.

## REFERENCES
1. H. Guo, C. Herkommer, A. Billat, D. Grassani, C. Zhang, M. H. P. Pfeiffer, W. Weng, C.-S. Brès, and T. J. Kippenberg, Nat. Photonics **12**, 330-335 (2018).
2. L. G. Wright, D. N. Christodoulides, and F. W. Wise, Science **358**, 94-97 (2017).
3. D. Richardson, J. Fini, and L. Nelson, Nat. Photonics **7**, 354 (2013).
4. G. Li, N. Bai, N. Zhao, and C. Xia, Adv. Opt. Photonics **6**, 413-487 (2014).
5. F. Poletti, and P. Horak, J. Opt. Soc. Am. B **25**, 1645-1654 (2008).
6. T. Herr, V. Brasch, J. D. Jost, C. Y. Wang, N. M. Kondratiev, M. L. Gorodetsky, and T. J. Kippenberg, Nat. Photonics **8**, 145 (2013).
7. L. G. Wright, D. N. Christodoulides, and F. W. Wise, Nat. Photonics **9**, 306-310 (2015).
8. G. Lopez-Galmiche, Z. Sanjabi Eznaveh, M. A. Eftekhar, J. Antonio Lopez, L. G. Wright, F. Wise, D. Christodoulides, and R. Amezcua Correa, Opt. Lett. **41**, 2553-2556 (2016).
9. K. Krupa, C. Louot, V. Couderc, M. Fabert, R. Guenard, B. M. Shalaby, A. Tonello, D. Pagnoux, P. Leproux, A. Bendahmane, R. Dupiol, G. Millot, and S. Wabnitz, Opt. Lett. **41**, 5785-5788 (2016).
10. S. Longhi, Opt. Lett. **28**, 2363-2365 (2003).
11. K. Krupa, A. Tonello, A. Barthélémy, V. Couderc, B. M. Shalaby, A. Bendahmane, G. Millot, and S. Wabnitz, Phys. Rev. Lett. **116**, 183901 (2016).
12. H. E. Lopez-Aviles, F. O. Wu, Z. S. Eznaveh, M. A. Eftekhar, F. Wise, R. A. Correa, and D. N. Christodoulides, APL Photonics **4**, 022803 (2019).
13. Z. Liu, L. G. Wright, D. N. Christodoulides, and F. W. Wise, Opt. Lett. **41**, 3675-3678 (2016).
14. K. Krupa, A. Tonello, B. M. Shalaby, M. Fabert, A. Barthélémy, G. Millot, S. Wabnitz, and V. Couderc, Nat. Photonics **11**, 237 (2017).
15. F. O. Wu, A. U. Hassan, and D. N. Christodoulides, Nat. Photonics, in press.
16. A. Picozzi, J. Garnier, T. Hansson, P. Suret, S. Randoux, G. Millot, and D. N. Christodoulides, Phys. Rep. **542**, 1-132 (2014).
17. S. Dyachenko, A. C. Newell, A. Pushkarev, and V. E. Zakharov, Physica D **57**, 96-160 (1992).
18. A. Yariv, Y. Xu, R. K. Lee, and A. Scherer, Opt. Lett. **24**, 711-713 (1999).
19. D. N. Christodoulides, F. Lederer, and Y. Silberberg, Nature **424**, 817 (2003).
20. A. Picozzi, Opt. Express **15**, 9063-9083 (2007).
21. T. M. Apostol, *Mathematical analysis* (Addison-Wesley, 1974).
22. S. Braun, J. P. Ronzheimer, M. Schreiber, S. S. Hodgman, T. Rom, I. Bloch, and U. Schneider, Science **339**, 52-55 (2013).